\documentclass[pra,aps,twocolumn,epsfig,showpacs]{revtex4}
\usepackage{amsmath}
\usepackage{amssymb}
\usepackage[dvips]{graphicx}%\usepackage{graphicx}% Include figure files
\usepackage{dcolumn}% Align table columns on decimal point
\usepackage{bm}% bold math
\usepackage[colorlinks=false,dvipdfm]{hyperref} % citecolor=blue �ļ��ڲ�����
\usepackage{setspace}
\usepackage{amsfonts}
\usepackage{rotating}
\usepackage{makeidx}
\usepackage{amsthm}

\newcommand{\bgeq}{\begin{eqnarray}}
\newcommand{\edeq}{\end{eqnarray}}

\begin{document}
%\begin{CJK*}{GB}{}

\title{Phase-Entanglement Complementarity with Time-Energy Uncertainty}

\author{Fu-Lin Zhang}
\email[]{flzhang@tju.edu.cn}
\affiliation{Department of Physics, School of Science, Tianjin University, Tianjin 300072, P. R. China}

\author{Mai-Lin Liang}
\affiliation{Department of Physics, School of Science, Tianjin University, Tianjin 300072, P. R. China}

\date{\today}

\begin{abstract}
We present a unified view of the Berry phase of a quantum system and its entanglement with surroundings.
The former reflects the nonseparability between a system and a classical environment as the latter for a quantum environment, and the concept of  geometric time-energy uncertainty can be adopted as a signature of the nonseparability.
Based on this viewpoint, we study their relationship in the quantum-classical transition of the environment, with the aid of a spin-half particle (qubit) model exposed to a quantum-classical hybrid field.
In the quantum-classical transition, the Berry phase has a similar connection with the time-energy uncertainty as the case with only a classical field,
whereas the geometric phase for the mixed state of the qubit exhibits a complementary relationship with the entanglement.
Namely, for a fixed time-energy uncertainty, the entanglement is gradually replaced by the mixed geometric phase as the quantum field vanishes.
And the mixed geometric phase becomes the Berry phase in the classical limit.
The same results can be draw out from a displaced harmonic oscillator model.
\end{abstract}


%\pacs{03.65.-w; 02.20.-a; 21.10.sf; 31.30.jx}
%\keywords{}

\maketitle
%\end{CJK*}

% \section{Introduction \label{intro}}

%{\bf PACS numbers:} 03.67.-a, 03.65.Ta

\section{Introduction}

%\emph{Introduction.--}
A quantum system always need to be isolated from its surroundings, which is referred to an open quantum system.
Otherwise, we would have to treat the system and \textit{the rest of the universe} \cite{Berry} as a whole.
An significant obstacle to this separation is the entanglement \cite{RevModPhys.81.865} between the open system and its environment caused by their quantum nature.
The system is exactly described by its reduced density matrix defined as the partial trace of a global state over the Hilbert space of environment.
In an ideal case, the decoherence of an open system due to entanglement with its environment is negligible.
The effects of the environment represents as parameters, or say classical fields, which are time-dependent in most cases, in the local Hamiltonian of the open system.
If the field is slowly altered, the open system will behave like a closed one, namely staying adiabatically in an instantaneous eigenstate of the time-dependent Hamiltonian.
The difference is that the open system undergoing a cyclic adiabatic evolution gains a Berry phase \cite{Berry}.
In Berry's original paper \cite{Berry}, the phase is said to be geometric because it results from the geometric properties of the parameter space of the local Hamiltonian.
In other words, it depends on the properties of the external field in evolutionary process, and thus reveals a nonseparable relation between the system and environment.

In the two aforementioned cases, the nonseparability between a system and an environment exhibits either entanglement or Berry phase as the environment is either quantum or classical.
This naturally leads to an interesting question: What is their relationship in the quantum-classical transition of the environment?
To answer this question, we introduce a model of a spin-half particle (qubit) coupled to an adiabatically rotating quantum-classical hybrid field.
And before that, pointed out that the Berry phase is caused by a neglected Hamiltonian and accompanied by a geometric time-energy uncertainty \cite{AAS90,Anandan91,Uhlmann92}.
This uncertainty is shown to be a signature of the total nonseparability between the system and the quantum-classical hybrid field.
Because of the entanglement between the principal qubit and the field, the original definition of the Berry phase is no longer applicable.
We study its two extensions, one of which is presented in this work based on the neglected Hamiltonian and the other is the geometric phase for mixed state \cite{Sj,Singh,Tong}.
In this work, we call the former the Berry phase and the latter the mixed geometric phase.
When the classical part of the field vanishes, one can find the Berry phase leads to the definition in the works of vacuum induced Berry phase \cite{Vacuum,VacuumM,VacuumRabi,VacuumRabi0} in the Jaynes-Cummings (JC) model \cite{JCM}, where they devise the phase with the aid of a phase shift operator.
In the quantum-classical transition of external field and for a fixed time-energy uncertainty, the mixed geometric phase replaces the entanglement gradually, and becomes the Berry phase in the classical limit.
This shows a complementary relationship between the mixed geometric phase and the entanglement to reflect the nonseparability.

%We begin from the case with only a classical field and show a simple relation between the Berry phase and the time-energy uncertainty.
%When the field is changed into a quantum-classical hybrid one, its entanglement with the system introduces a correction to the time-energy uncertainty.
%That is, the entanglement, the nonseparability between a system and a quantum environment, is accompanied by a time-energy uncertainty, as well as the Berry phase for the case of classical field.
%From the relation between the Berry phase and time-energy uncertainty, one can find that the classical part and quantum part of the external field are parallel to each other but behave as a classical field in another direction, where the deflection is caused by the quantum fluctuation.
%In particular, we show a complementary relationship between the mixed geometric phase and the entanglement to reflect the nonseparability.
%Namely, in the quantum-classical transition of external field and for a fixed time-energy uncertainty, the entanglement is gradually replaced by the mixed geometric phase, and the latter becomes the Berry phase in the classical limit.
%These results are verified in section \textit{Harmonic Oscillator} by using a displaced harmonic oscillator model.

\section{Neglected Hamiltonian}
%\emph{Neglected Hamiltonian.--}
Before introducing our model, we first carefully examine the connection between the Berry phase of an open system and its nonseparability with a classical environment.
The connection will become apparent if we consider the Berry phase as the adiabatic limit of the Aharonov-Anandan (AA) phase \cite{Aharonov,ZhengJY95,ZhengJY96}.
The latter is an extension of the Berry phase without adiabatic approximation.
It has been shown that only for a nonstationary state after a cyclic evolution might the AA phase appear \cite{ZhengJY95}.
That is, if an open quantum system stays in an instantaneous eigenstate $|\psi\rangle$ of its Hamiltonian $H_0$ and acquires an AA phase after a cyclic evolution, there should exist a Hamiltonian $H_{\Delta}$ which is neglected but affects the system in addition to $H_0$, and $[H_0,H_{\Delta}] \neq 0$ \cite{Berry09}.
Therefore, the uncertainty in energy when the system is isolated from its environment causes the AA phase.
In the adiabatic limit, $H_{\Delta} \rightarrow 0$, the expectation value of the Hamiltonian $H= H_0  + H_{\Delta}$ becomes the instantaneous eigenvalue of $H_0$, and the AA phase returns the Berry phase.
The Berry phase comes from the fact that even the uncertainty of energy tends to zero in the adiabatic limit, its effect in a cyclic evolution is finite and nonzero.
The accumulating result of the uncertanty of energy can be represented as the geometric quantum uncertainty relation \cite{AAS90,Anandan91,Uhlmann92}
\begin{eqnarray}\label{S}
\mathcal{S}=2  \int^{\tau}_0 \Delta E d t,
\end{eqnarray}
where $\Delta E$ is the energy fluctuation, defined by
\begin{eqnarray}\label{ED}
\Delta E^2 = \langle \psi | H^2 |\psi\rangle -\langle \psi | H |\psi\rangle^2.
\end{eqnarray}
It is the distance that the system traverses during its evolution in the projective Hilbert space measured by Fubini-Study metric \cite{AAS90}.
In the following section, we will extend the Berry phase to a system in a quantum-classical hybrid field with the aid of the neglected Hamiltonian, and show the connection between the time-energy uncertainty and its entanglement with the field.

\section{Qubit Model}

\begin{figure}%[!htbp]
%\centering
\centerline{\includegraphics[width=4cm]{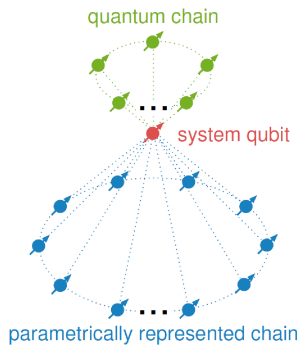}}
\caption{
The spin star model with two sets of environmental spins.
One of the sets is a quantum spin chain with a total spin $j$, and the other is treated at a classical level as an external field $\mathbf{B}$.
\label{fig1} }
\end{figure}

%\emph{Qubit Model.--}
Let us now consider the model of a qubit coupled to a rotating quantum-classical hybrid field with the Hamiltonian
\begin{eqnarray}\label{H0}
 H_0=  - \mu  \mathbf{J}  \cdot {\bm  \sigma} - \mathbf{B} \cdot  {\bm  \sigma}
\end{eqnarray}
where ${\bm  \sigma}=(\sigma_x, \sigma_y, \sigma_z)$ are the Pauli
operators of the qubit, $\mathbf{B} = B(\sin\theta \cos \omega t,
\sin\theta \sin \omega t,\cos \theta)$ is the classical part of the
field, and $\mu  \mathbf{J}$ denotes the quantum part with $
\mathbf{J}=(J_x, J_y, J_z)$ being the angular momentum operators
of a spin-$j$ particle. Experimental realization of the interaction
between two spins and observation of the Berry phase for such a
system is feasible by current technology \cite{ExNat,ExDu}. In the
frame of quantum optics, the physical meaning of the pure quantum
term can be easily understood if we consider the spin operator
$\mathbf{J}$ as the Schwinger representation \cite{Vacuum}
of two modes of a quantized optical field.
The Hamiltonian $ - \mu  \mathbf{J}  \cdot {\bm  \sigma}$ denotes a
interacting process between the two modes of optical field and a two
level system conserving the total photon number. The whole
Hamiltonian \eqref{H0} could also be regarded as a
semiclassical spin star model
\cite{PNAS,IJTP,SpinStar1,SpinStar2}. As shown in Fig. \ref{fig1}, the system
qubit interacts with two Heisenberg chains, i.e. two sets of
environmental spins. One of the chains has a total spin $j$, and the other is treated as a classical field $\mathbf{B}$ as its entanglement with the system qubit is negligible \cite{PNAS,IJTP}.
 The quantum-classical transition of the environment can be shown by tuning the parameter $\mu j /B$ from infinity to zero.
Thus the experimental realization of the Hamiltonian \eqref{H0} provides a simulation of the crossover from a quantum to a classical environment.

%\subsection{Classical Field}
\emph{Classical Field.--}
 For the simplest case in the absence of quantum field, where $\mu=0$, we can treat the principal qubit in
the space of itself. The instantaneous eigenstates of $H_0$
corresponding to the eigenvalues $\varepsilon_{\pm}=\mp B$ are
\begin{eqnarray} \label{psipm}
| \psi^{+} \rangle = U(t)|\uparrow\rangle,\ \ \ | \psi^{-} \rangle = U(t)|\downarrow\rangle,
\end{eqnarray}
with $U(t)= \exp(- i \omega t \sigma_z /2) \exp(- i \theta \sigma_y
/2) $. To keep the system in the eigenstates $| \Psi^{\pm}
\rangle=c_{\pm} | \psi^{\pm} \rangle$, the neglected Hamiltonian
should be $H_{\Delta}=i \dot{U}(t) U^{\dag}(t)=\omega \sigma_z /2$,
and the coefficients $c_{\pm}=\exp (-i \varepsilon_{\pm} t)$.
 After a period $\tau=2 \pi /\omega$, the total phases
$
\phi_{\pm} = - \varepsilon_{\pm} \tau \mp \pi
$
are determined by the eigenvalues and the property of the spin-half rotation operation.
For the dynamic phase, which is defined by $\phi^d = -   \int^{\tau}_0 \langle   H    \rangle d t$ \cite{Aharonov},  we get
$
\phi^d_{\pm} = - \varepsilon_{\pm} {\tau}  \mp  \pi \cos \theta.
$
Hence the Berry phase, as the AA phase $\gamma=\phi-\phi^d$ in the limit of $\omega \rightarrow 0$,  can be written as
\begin{eqnarray}\label{Cphase}
\gamma_{\pm} = \mp \pi (1-\cos \theta).
\end{eqnarray}
They equal the half of the solid angle $\Omega$ subtended by the
path followed by $\mathbf{B}$ in the parameter space,
 and the $\mp$ sign depends on whether the spin was aligned or against the direction of the field.
Although the fluctuation of energy defined by \eqref{ED} approaches
zero in the adiabatic limit, the time-energy uncertainty \eqref{S}
in a period is a constant independent of the frequency $\omega$ as
\bgeq \mathcal{S}_{\pm}= 2 \pi \sin \theta. \edeq
which is nothing
but the shortest perimeter surrounding the solid angle $\Omega$ in
an unit sphere. In this work, we confine ourselves to the case of
rotating fields not only for simplicity but also to gain as large
the geometric phases as possible under a fixed time-energy uncertainty.

 Furthermore, if we consider the geometric phase for noncyclic evolution \cite{Samuel} and require $|\gamma|=\pi$ corresponding to a reversal of the interference fringes, we can easily obtain a time-energy uncertainty relation in the present model
 \cite{note1}
\bgeq \label{DEDT} \langle \Delta E \rangle \Delta t \geq
\frac{h}{2}, \edeq where $\langle \Delta E  \rangle$ is the
time-averaged uncertainty in energy during the time $\Delta t$ to
gain the geometric phase  $|\gamma|=\pi$ and $h$ is the Planck
constant. Equality in \eqref{DEDT} holds when the classical field is
perpendicular to z axis, i.e. $\theta=\pi/2$.

%\subsection{Quantum-Classical Hybrid Field}
\emph{Quantum-Classical Hybrid Field.--}
We will now derive the results for the full Hamiltonian \eqref{H0}.
One can diagonalize the Hamiltonian in the subspace of
$\{\mathcal{U}(t)|m\rangle|\uparrow\rangle,
\mathcal{U}(t)|m+1\rangle|\downarrow\rangle\}$ , where $|m\rangle$
is the eigenvector of $J_z$ with $J_z |m\rangle= m |m\rangle$ and
$\mathcal{U}(t)=\exp[- i \omega t(-j+J_z+ \sigma_z /2)] \exp[- i
\theta (J_y + \sigma_y /2)].$ In the unitary transformation, a term
$-j\omega t$ is introduced to eliminate the influence of the $\pi$
phase brought by a odd $j$ \cite{note2}.
The instantaneous eigenstates of $H_0$ corresponding to eigenvalues $\varepsilon^{\pm}_m$ are
\begin{eqnarray}\label{psim}
| \psi^{+}_m \rangle = \mathcal{U}(t) \left(\cos \frac{\alpha_m}{2} |m\rangle|\uparrow\rangle + \sin \frac{\alpha_m}{2} |m+1\rangle|\downarrow\rangle \right),&& \nonumber \\
| \psi^{-}_m \rangle = \mathcal{U}(t) \left(\sin \frac{\alpha_m}{2} |m\rangle|\uparrow\rangle - \cos \frac{\alpha_m}{2} |m+1\rangle|\downarrow\rangle \right),&&
\end{eqnarray}
with $\tan \alpha_m = \mu \sqrt{j(j+1)-m(m+1)}/[B+\mu(m+1/2)]$.
Here, the values of index $m$ range from $-(j+1)$ to $j$. For $m=j$
or $-(j+1)$, only two states $| \psi^{+}_j\rangle = \mathcal{U}(t)
|j\rangle|\uparrow\rangle$ and $| \psi^{-}_{-(j+1)}\rangle = -
\mathcal{U}(t)|-j\rangle|\downarrow\rangle$ are physically possible.
They are direct products of two spin-coherent states aligned or
against the classical field, and
the states of spin-$j$ are the closest to the classical states
\cite{SpinAA}. The concurrence \cite{Wootters} of states \eqref{psim}
 can be written as \bgeq \mathcal{C}= \sin \alpha_m
\edeq which is the degree of entanglement between the system qubit
and its external field.

In the case of the classical field discussed above, the dynamics of the field $\mathbf{B}$ is not described by the Hamiltonian \eqref{H0}, but is considered as a time-dependent variable.
A quantum field cannot be treated in this way anymore, as its state can be influenced by the interaction with the principal qubit.
We have to take into account a Hamiltonian driving the quantum field to rotate with the same frequency as the classical field.
The most natural choice of the Hamiltonian is $H_j= (-j+J_z)\omega$.
 Staying in the instantaneous eigenstates \eqref{psim} requires the whole system to satisfy the Schr\"{o}dinger equation
\bgeq i | \dot{\Psi}^{\pm}_m \rangle=( H_0 +H_j+ H_{\Delta}) |
\Psi^{\pm}_m \rangle, \edeq with $| \Psi^{\pm}_m\rangle= c^{\pm}_m |
\psi^{\pm}_m \rangle$. Consequently, the neglected Hamiltonian can
be found as $H_{\Delta}=i \dot{\mathcal{U}}(t)
\mathcal{U}^{\dag}(t)-H_j=\omega \sigma_z /2$, and the coefficients
$c^{\pm}_m=\exp (-i \varepsilon^{\pm}_m t)$. After a period, the
total phase is $ \phi_{\pm} = - \varepsilon^{\pm}_m \tau \mp \pi, $ and
the dynamic phase is $\phi^d_{\pm} = -   \int^{\tau}_0 \langle
\Psi^{\pm}_m|  H |\Psi^{\pm}_m   \rangle d t  = -
\varepsilon^{\pm}_m {\tau}  \mp  \pi \cos \theta \cos \alpha_m $
with $H=H_0+H_{\Delta}$. Hence we obtain the Berry phase
\begin{eqnarray}\label{Qphase}
\gamma^q &=&   \pi (1-\cos \theta \cos \alpha_m) \nonumber \\
&=&   \pi (1-\cos \theta) +     \pi(1- \cos \alpha_m) \cos\theta,
\end{eqnarray}
where we omit the $\pm$ sign for clarity. The second term in
\eqref{Qphase} shows a correction in Berry phase aroused by the
quantum part of the field.
 We call this term quantum field induced Berry phase (QBP), as it is caused by quantum fluctuation and entanglement.
In addition, when the classical field vanishes, one can choose $\theta=0$ due to the symmetry of the Hamiltonian \eqref{H0} and find the QBP has the same physical meaning as the vacuum induced Berry phase \cite{Vacuum,VacuumM,VacuumRabi,VacuumRabi0}.
  A remarkable quantum nature of
QBP can be understood by its sharp contrast to the classical one.
Namely, the Berry phase is robust against the perturbations by
classical environment, particularly is invariant under a changing of
the strength of the field. But it is sensitive to the quantum
perturbation even though the polarization direction of the spin-$j$
particle in the eigenstates remains parallel to the classical field.
From the form of QBP in \eqref{Qphase} one can surmise it revelent
to a solid angle in a space corresponding to the quantum fluctuation
of spin-$j$. The QBP reaches its maximum for the most entangled
states with $m=0$ or $-1/2$ and vanishes for the two separable
eigenstates $| \psi^{+}_j\rangle$ and $| \psi^{-}_{-(j+1)}\rangle$.
And in the former case the states give rise to the maximum
dispersion for square $\mathbf{J}^2$ of the spin angular momentum, and the
latter two states lead to the minimum \cite{SpinAA}.

For further analysis of the physical meaning of QBP and the whole
Berry phase in \eqref{Qphase} we calculate the geometric
time-uncertainty relation \eqref{S} during a period of rotation
$\tau=2\pi/\omega$. Substituting the expressions of $| \Psi^{\pm}_m
\rangle$ and $H$ into \eqref{S} leads to
\bgeq \mathcal{S}^q = 2 \pi \sqrt{1- \cos^2 \theta \cos^2 \alpha_m}.
\edeq
Similar to QBP, it is increased by the entanglement between
the qubit and field. It is interesting to note that if one defines
an angle $\cos \theta_m = \cos  \theta \cos  \alpha_m$, there also
exists a relationship of solid angle and perimeter between
$\gamma^q$ and $\mathcal{S}^q$ as the case of the classical field.
That is, the classical part and quantum part of the external field
parallel to each other behave as an effective classical field in
another direction. This deflection caused by the fluctuation of the
quantum field is most visible when $\theta =0$, where only the
quantum field contributes to $\mathcal{S}^q$ and $\gamma^q$, and
vanishes for $\theta =\pi/2$ with the classical filed making the
most contribution.
 Moreover, for a fixed value of geometric phase $\gamma^q=\pi$, the inequality \eqref{DEDT} still holds for the qubit system with a quantum-classical hybrid field.
And the entanglement between the principal qubit and its external field reduces the value of $\langle \Delta E \rangle \Delta t $ for a given $\theta$.

In the above discussion about QBP, we treat the system qubit and the
quantum field as a composite system when we consider their
evolution. We actually remove terms irrelevant to the qubit in the
total phase of the whole system and obtain the Berry phase \eqref{Qphase}.
That is, the Berry phase \eqref{Qphase} depends on the evolution of
the quantum field entangled with the qubit. To study the phase
determined only by the geometry of the path of the qubit, we derive
the time-dependent mixed state \bgeq \rho= \sin^2 \frac{\alpha_m}{2}
|\psi^+\rangle \langle \psi^+|+\cos^2 \frac{\alpha_m}{2}
|\psi^-\rangle \langle \psi^-|, \edeq by tracing out the quantum
filed in the eigenstates \eqref{psim}, where
$ |\psi^{\pm}\rangle$ is defined in \eqref{psipm}. Here, we only
give the case for the pure state $| \psi^{-}_m \rangle$, the mixed
geometric phase for $| \psi^{+}_m \rangle$ can be easily obtained by
changing $\alpha_m$ into $\pi + \alpha_m$. We can calculate the
mixed geometric phase by using the definition with a kinematic
description \cite{Sj,Singh,Tong} and get \bgeq\label{Mphase}
\gamma^{mix} &=& \arg \left(\sin^2 \frac{\alpha_m}{2} e^{i \gamma_+ }+ \cos^2 \frac{\alpha_m}{2} e^{i \gamma_-}\right) \nonumber \\
&=& \arctan \left(\cos\alpha_m \tan \gamma_-  \right),
%&=& \arctan \left\{ \tan [\pi(1-\cos\theta)] \cos\alpha_m \right\}
\edeq where $\gamma_{\pm}$ are the Berry phases \eqref{Cphase}
resulting from the classical field. The phase $\gamma^{mix}$ is
manifestly gauge invariant and can be experimentally tested in
interferometry.

Let us give a further discussion about the relationship among these
quantities related to the nonseparability between the system and the
external field. For a fixed value of $\theta$, $\gamma^{mix} \leq
\gamma^q$, and the entanglement widens their gaps. They are equal to
the Berry phase $\gamma_{-}$ for a separable eigenstate. It is
important to note that the time-energy uncertainty leads to an upper
bound of the entanglement $\mathcal{C} \leq \mathcal{S}^q/ (2 \pi)$.
The equality holds when the mixed geometric phase $\gamma^{mix}$
vanishes. These reveal a complementary relationship between the
mixed geometric phase $\gamma^{mix}$ and the entanglement to reflect
the nonseparability between the system and the external field, while
$\mathcal{S}^q$ and $\gamma^q$ can be considered to be their
sum. For a fixed amount of $\mathcal{S}^q$,
the entanglement is gradually replaced by the geometric phase
$\gamma^{mix}$ in the quantum-classical transition of external
field, and becomes the Berry phase when the quantum field
vanishes.
In a figurative sense, \emph{Berry phase is semiclassical entanglement between a quantum system and a classical environment.}

\section{Harmonic Oscillator}
%\emph{Harmonic Oscillator.--}
 We also study a displaced harmonic
oscillator which is another canonical example of the Berry phase
\cite{HO} to verify our above discussions. Here we add a qubit as
the part of the its environment, which is described in terms of the
Pauli operators $\sigma_z, \sigma_{\pm}=(\sigma_x \pm i
\sigma_y)/2$. Suppose the qubit-oscillator interaction is
characterized by a JC term \cite{JCM}, we
get \bgeq H_0= \nu b^{\dag} b + g (\sigma_- b^{\dag} + \sigma_+ b ),
\edeq where $ b^{\dag}=a^{\dag}-\beta^*, b= a - \beta$ are displaced
creation and annihilation operators with the frequency $\nu$,  $g$
is the coupling constant. We take $\beta= |\beta| e^{-i \omega t}$ and
$\beta^*= |\beta| e^{i \omega t}$ to be slowly rotating parameters.
One can find its instantaneous eigenstates in the subspace $\{
\mathcal{V}(t)|n\rangle|\uparrow\rangle,\mathcal{V}(t)|n+1\rangle|\downarrow\rangle
\}$, where $\mathcal{V}(t)=\exp(\beta b^{\dag} - \beta^* b) \exp[- i
\omega t (\sigma_z + 1)/2]$, and $|n\rangle$ satisfies $a^{\dag}a|n\rangle=n|n\rangle$. The
instantaneous eigenstates with eigenvalues $\varepsilon^{\pm}_n$ are
derived as
\begin{eqnarray}\label{psimh}
| \psi^{+}_n \rangle = \mathcal{V}(t) \left(\cos \frac{\alpha_n}{2} |n\rangle|\uparrow\rangle - \sin \frac{\alpha_n}{2} |n+1\rangle|\downarrow\rangle \right),&& \nonumber \\
| \psi^{-}_n \rangle = \mathcal{V}(t) \left(\sin \frac{\alpha_n}{2} |n\rangle|\uparrow\rangle + \cos \frac{\alpha_n}{2} |n+1\rangle|\downarrow\rangle \right),&&
\end{eqnarray}
where $\tan \alpha_n = g \sqrt{n+1}/\nu$. The concurrence of states
$| \psi^{\pm}_n \rangle$ is
\bgeq
\mathcal{C}_h= \sin \alpha_n ,
\edeq
which is zero for the state $|0\rangle|\downarrow\rangle$, corresponding to the eigenstate $| \psi^{-}_n \rangle$ in \eqref{psimh} with $n=-1$.
 We also choose a
Hamiltonian to drive the rotation of the qubit with the frequency of
$\beta$, which is $H_q = \omega(\sigma_z + 1)/2 $.
The neglected Hamiltonian in this case is $H_{\Delta}=\omega
a^{\dag} a$. Follow the same steps of the qubit case in the above
section, we obtain the time-energy uncertainty \bgeq
\mathcal{S}^q_h= 2 \pi \sqrt{4(2n+1)|\beta|^2 + \sin^2 \alpha_n},
\edeq the Berry phase \bgeq \gamma^q_h=  2 \pi |\beta|^2  \pm
\pi(1-\cos \alpha_n), \edeq and the mixed geometric phase \bgeq
\gamma^{mix}_h=  2 \pi |\beta|^2. \edeq Here, both the phases are
defined modulo $2\pi$.
 Obviously there exist an entanglement induced term in each of the expression of the time-energy uncertainty and the Berry phase.
And the entanglement and the  mixed geometric phase are complementary for a fixed value of $\mathcal{S}^q_h$.

\section{Summary}

%For instance, the vacuum induced Berry phase \cite{Vacuum} mentioned above has brought about a series of studies for more interesting physical interpretation and application \cite{VacuumM,VacuumRabi,VacuumRabi0}, and the very recent parametric representation of an open systems proposes an approach to studying its behavior along the quantum-classical transition of its environment.

%\emph{Summary.--}
In this work, we present a viewpoint of the Berry phase that it reflects the nonseparability between an open system and its classical environment just like the entanglement between a system and a quantum environment.
This unified view of the two concepts inspires us to explore their properties and connection in the quantum-classical transition of the environment.
The viewpoint is supported by the fact that the Berry phase can be considered as the effect of a neglected Hamiltonian which affects the system but has no effect on the eigenvalues and eigenstates in the
adiabatic limit.
This understanding can be obtained by checking the relation between the Berry phase and the AA phase, and provides an approach to extend the Berry phase to the system in a quantum-classical hybird environment.
In addition, the geometric time-energy uncertainty as an accumulating result of the neglected Hamiltonian in a cyclic evolution is found to be a signature of the nonseparability not only for a classical environment but also for a quantum-classical hybird one.

Based on foregoing considerations we investigate a qubit under an adiabatically rotating quantum-classical hybrid field.
The entanglement between the principal qubit and the field introduces a correction to the time-energy uncertainty and a corresponding effect on the Berry phase.
For a fixed time-energy uncertainty, the entanglement is gradually replaced by the mixed geometric phase, determined solely by the geometry of the the qubit, in the quantum-classical transition of external field.
That is, the geometric phase and entanglement have complementary relationship to reflect the nonseparability between the system and the external field.
We also make similar calculations for a model of displaced harmonic oscillator to verify these conclusions, in which a qubit acts as the quantum part of the field.

%Berry phase and quantum entanglement are two fundamental concepts in quantum theory, both of which have applications in the realm of quantum information theory and computation \cite{Nielsen,zanardi99,sj2012}.
%Their relation and has extensively studied from various angles and many interesting results or methods have been reported \cite{Sj1,Hessmo,Tong1,Tong2,Williamson,yi1,Li,EPL},
%such as the vacuum induced Berry phase \cite{Vacuum,VacuumM,VacuumRabi,VacuumRabi0} mentioned above and parametric representation of an open systems \cite{PNAS,IJTP} proposed very recently.
We have been very careful to introduce a quantum part of the environment in our models without giving up the conditions of adiabaticity and cyclicity in Berry's original definition of geometric phase.
These models allow us to treat the geometric phases and quantum entanglement uniformly. %, which reflect the nonseparability between an open system and its classical and quantum environments respectively.
We believe that, by removing the restrictions in this work, one may uncover more general connections among geometric aspects of quantum mechanics and different quantum correlations \cite{UnifDiscord}.

\begin{acknowledgments}
The idea was initiated in our discussions with Wu-Sheng Dai, Mi Xie and Da-Bao Yang.
We are very indebted to Sir Michael Berry for his remarks and encouragement.
F.L.Z. also thanks Jing-Ling Chen, Mang Feng, Paola Verrucchi, and Kamal Berrada for their valuable comments.
This work was supported by NSF of China (Grant No. 11105097).
\end{acknowledgments}

\bibliography{EntanglementBerryPhase}

\begin{thebibliography}{30}
\bibitem[{not({\natexlab{a}})}]{note1}
\bibinfo{note}{In the full text but this equation we set the reduced Planck
  constant $\hbar=1$.}
\bibitem[{not({\natexlab{b}})}]{note2}
\bibinfo{note}{This shows the Berry phases induced by quantum field we obtained
  below violate the $U(1)$ gauge invariance, as it can be affected by the
  choice of instantaneous eigenstates. However the difference of the Berry
  phases between two eigenstates is an invariant under the $U(1)$ gauge
  transformations. This remark is also applicable to the results in
  \cite{Vacuum} as well. One can change the Berry phases in \cite{Vacuum} by
  replacing the phase shift operator by $U(\phi)= \exp \{-i \phi [a^{\dag}a +
  \delta (\phi)]\}$ without influence on the rotation of the field.}
\end{thebibliography}

\end{document}